\documentclass[11pt,a4paper]{article}
 \pdfoutput=1

\usepackage{jheppub_kim}
\usepackage{rotating} 
\usepackage{graphicx,epsfig}
\usepackage{amsmath}
\usepackage {amssymb}
\usepackage{subfigure}

\usepackage{relsize}

\renewcommand{\(}{\left(}
\renewcommand{\)}{\right)}

\def\L{\mathcal{L}}
\def\K{\mathcal{K}}
\def\X{\mathcal{X}}

\def\d{\mathrm{d}}
\def\mn{_{\mu \nu}}

\def\mupn{^\mu_{\phantom{\mu}\nu}}
\def\({\left(}
\def\){\right)}

\def\p{\partial}

\makeatletter

\providecommand{\U}[1]{\protect\rule{.1in}{.1in}}

\newcommand{\be}{\begin{equation}}
\newcommand{\ee}{\end{equation}}
\newcommand{\ba}{\begin{eqnarray}}
\newcommand{\ea}{\end{eqnarray}}

\usepackage{hyperref}        
\hypersetup{colorlinks=true,urlcolor=blue}

\usepackage[misc]{ifsym}

\title{Observational constraints on   extended Proca-Nuevo gravity and 
cosmology}

\author[a]{Fotios K. Anagnostopoulos}

\author[b,c,d]{Emmanuel N. Saridakis}

\affiliation[a]{Department of Informatics and Telecommunications, 
University of Peloponnese, Karaiskaki 70, 22100, Tripoli, Greece}

\affiliation[b]{National Observatory of Athens, Lofos Nymfon, 11852 Athens, 
Greece}

\affiliation[c]{CAS Key Laboratory for Research in Galaxies and Cosmology,  
 University of Science and Technology of China, Hefei, Anhui 230026, China}

 \affiliation[d]{Departamento de Matem\'{a}ticas, Universidad Cat\'{o}lica del 
Norte, Avda.
Angamos 0610, Casilla 1280 Antofagasta, Chile}

\abstract{  We confront massive Proca-Nuevo  gravity with cosmological 
observations. The former is a  non-linear theory  involving a massive spin-1 
field, that can be extended incorporating operators of the Generalized Proca 
class, and when coupled to gravity it can be covariantized in a way that 
exhibits consistent and ghost-free cosmological solutions, without experiencing 
instabilities and superluminalities at the perturbative level. 
When applied at a cosmological framework it induces extra terms in the 
Friedmann equations, however due to the special non-linear construction the 
field is eliminated in favor of the Hubble function. Thus, the resulting 
effective dark energy sector is dynamical,  however it contains
 the same number of free parameters with the  $\Lambda$CDM concordance model.  
We use data from Supernovae Ia (SNIa) and Cosmic Chronometers 
(CC)  observations and we construct the corresponding likelihood-contours for 
the free parameters. Interestingly enough, application of various information 
criteria, such as AIC, BIC and DIC, shows that the scenario of  massive 
Proca-Nuevo  gravity, although having exactly the same number of free 
parameters 
with $\Lambda$CDM paradigm, it is  more efficient in fitting the data. 
Finally, the  reconstructed    dark-energy  equation-of-state parameter  
shows statistical compatibility  
with the model-independent, data-driven reconstructed one.
 }


\begin{document}
\maketitle

\section{Introduction}
 
 The concordance model of cosmology, namely  $\Lambda$CDM paradigm, which is 
based on general relativity with a cosmological constant, on the standard-model 
particles, and on cold dark matter, has been proven very efficient in 
describing 
the Universe evolution both at the background and perturbation levels. 
Nevertheless, according to recent observations of various origins, 
 $\Lambda$CDM  predictions seem to be in   tension  with the data, as for 
instance  the $H_0$ tension \cite{DiValentino:2020zio}, the 
$\sigma_8$ tensions \cite{DiValentino:2020vvd} etc (for reviews see 
\cite{Perivolaropoulos:2021jda,Abdalla:2022yfr}). On the other hand, at the 
theoretical level, $\Lambda$CDM faces the cosmological constant problem 
\cite{Weinberg:1988cp,Sahni:1999gb}, while  general relativity itself is 
non-renormalizable and thus it cannot be brought close to a quantum description 
\cite{Addazi:2021xuf,AlvesBatista:2023wqm}. Hence, a large amount of research 
has been devoted in constructing gravitational modifications, namely theories 
that 
possess general relativity as a limit but presenting theoretical as well as 
phenomenological advantages (for a review see  \cite{CANTATA:2021ktz}).

A first subset of modified gravity theories emerges by enhancing the 
Einstein-Hilbert Lagrangian through the inclusion of supplementary terms. This 
results in a diverse array of formulations, encompassing $f(R)$ gravity 
\cite{Starobinsky:1980te,DeFelice:2010aj}, $f(G)$ gravity \cite{Nojiri:2005jg}, 
$f(P)$ gravity \cite{Erices:2019mkd}, and Lovelock gravity 
\cite{Lovelock:1971yv}. Nonetheless, an alternative trajectory involves 
transcending the conventional curvature-centric approach to gravity by 
integrating other geometrical parameters, such as torsion and non-metricity. 
More specifically, one can embark from the teleparallel equivalent of general 
relativity \cite{Aldrovandi:2013wha,Maluf:2013gaa}, utilizing the torsion 
scalar 
$T$ 
as a Lagrangian and extending it to $f(T)$ gravity 
\cite{Bengochea:2008gz,Cai:2015emx}. Alternatively, one may extend the scope by 
adopting the symmetric teleparallel theory, utilizing the non-metricity scalar 
as the Lagrangian \cite{Nester:1998mp}, and further extending it to $f(Q)$ 
gravity \cite{BeltranJimenez:2017tkd,Heisenberg:2023lru}. All these 
classifications of gravitational theories demonstrate rich cosmological 
behaviors, captivating the keen interest of the scientific community
\cite{Cognola:2006eg, Nojiri:2010wj,Cai:2009zp, Chen:2010va,Tsujikawa:2009ku,
Harko:2011kv,Saridakis:2008fy,Bamba:2012cp, Clifton:2011jh, 
Skugoreva:2014ena,Darabi:2014dla,Kofinas:2014owa,Krssak:2015oua,
Basilakos:2016xob,Nunes:2016drj,Vagnozzi:2019ezj,Basilakos:2019mpe,
Yan:2019gbw,Pan:2020bur,Basilakos:2020qmu,Braglia:2020auw,Ilyas:2020zcb,
Anagnostopoulos:2021ydo,Benisty:2021cin, 
Capozziello:2022wgl,Chatzifotis:2022ubq,Cai:2023ykr,
Basilakos:2023xof,Gomez-Valent:2023hov,Basilakos:2023jvp,Millano:2023gkt}.  

One interesting subclass of modified gravity arises by considering the graviton 
to be massive. The inquiry into whether the graviton can possess mass has 
intrigued theorists for an extensive period. Originating with Fierz and Pauli 
\cite{Fierz:1939ix}, the subsequent development of a comprehensive nonlinear 
framework for massive gravity encountered a significant challenge, i.e. the 
emergence of the Boulware-Deser (BD) ghost \cite{Boulware:1973my}. This 
inherent 
problem persisted for decades until a notable breakthrough, marked by the 
introduction of a specific nonlinear extension of massive gravity by de Rham, 
Gabadadze, and Tolley (dRGT) \cite{deRham:2010ik}. Through meticulous 
Hamiltonian constraint analysis \cite{Hassan:2011hr} and an effective field 
theory approach \cite{deRham:2011rn}, it was demonstrated that the BD ghost 
could be eradicated by introducing a secondary Hamiltonian constraint (see  
\cite{Hinterbichler:2011tt} for a review). Beyond its theoretical 
significance, this nonlinear construction offers an additional advantage by 
potentially explaining the observed late-time cosmic acceleration: since by    
fine-tuning the graviton mass to an adequately small value, gravity 
becomes weaker at cosmological scales, then  the graviton potential can 
effectively emulate a cosmological constant \cite{deRham:2010tw, 
Gumrukcuoglu:2011ew}. However, the basic versions of the theory exhibit 
instabilities at the perturbative level \cite{DeFelice:2012mx}.

 Inspired by the structure of massive gravity, the 
generalized Proca action for a vector field with derivative self-interactions, 
resulting in a theory with only three propagating degrees of freedom, was 
proposed in \cite{Allys:2015sht,DeFelice:2016yws}. In particular, a consistent, 
local theory of a massive vector field without the presence of ghost-like 
instabilities was constructed, and thus the cosmological 
solutions at  both background and perturbation levels 
  were studied \cite{DeFelice:2016yws}. The generalized Proca   theory 
proves to have interesting cosmological phenomenology 
\cite{Heisenberg:2016eld,DeFelice:2016uil, Amado:2016ugk,DeFelice:2016cri, 
BeltranJimenez:2016afo, Minamitsuji:2016ydr, deFelice:2017paw, 
Nakamura:2017dnf, Domenech:2018vqj, BeltranJimenez:2019wrd, 
GallegoCadavid:2019zke, DeFelice:2020sdq, Heisenberg:2020jtr, 
  Geng:2021jso, ErrastiDiez:2021ykk, Dong:2023xyb}.  

Recently, an extended Proca 
theory, namely Proca-Nuevo (PN) theory, appeared in the literature 
\cite{deRham:2020yet}. It corresponds to 
a non-linear theory of a massive spin-1 field, that can be extended by 
incorporating operators of the Generalized Proca class without compromising the 
essential primary constraint crucial for consistency. When  the theory is 
combined with gravity it can be covariantized in models that support consistent 
and ghost-free cosmological solutions. In particular, 
 they  exhibit hot Big 
Bang solutions featuring a late-time self-accelerating epoch, and additionally 
at the perturbative level  certain 
sub-classes of the theory  satisfy all stability and subluminality 
conditions, and intriguingly, gravitational waves propagate at the speed of 
light.  Moreover, further cosmological solutions of the theory have been 
studied in \cite{deRham:2021efp}, while  the complete analysis of the 
constraint algebra has been performed in 
\cite{deRham:2023brw,ErrastiDiez:2023gme}. Finally, 
  the quantum stability of Proca-Nuevo interactions was 
explored in \cite{deRham:2021yhr}, where it was found that Proca-Nuevo   
and   generalized Proca theories have analogous behaviour at the quantum level, 
opening the door to speculations for being specific cases of a more general 
theory. 

In the present work we desire to confront massive Proca-Nuevo gravity and
 cosmology  with 
cosmological data from Supernovae Ia (SNIa) and Cosmic Chronometers (CC) 
observations, and extract constraints on the model parameters. The plan of the 
work is the following: In Section \ref{Sec1}   we present extended Proca-Nuevo  
 gravity and we apply it at a cosmological framework. In Section  \ref{Sec3} we 
provide the  datasets that we will use in our 
analysis,  and we analyze the various information criteria used for model 
comparison. Then, in Section \ref{Sec4} we perform the observational 
confrontation and we present the results. Finally, Section \ref{Conclusions} is 
devoted to the Conclusions.

\section{Extended Proca-Nuevo gravity and cosmology}
\label{Sec1}

In this section we  briefly present extended Proca-Nuevo theory (EPN) coupled 
 with gravity, and we apply it at a cosmological framework, following  
\cite{deRham:2021efp}.
 The action of the covariant  extended Proca-Nuevo theory is based on 
  a Lorentz massive spin-1 field, and it is written as
\begin{equation}
    \label{action}
    \mathcal{S} = \int d^4x\sqrt{-g}\left(\frac{M_{Pl}^2}{2}R + 
\mathcal{L}_{EPN} + \mathcal{L}_{M} \right),
\end{equation}
where   $R$ is the Ricci scalar, and $\L_M$ is   the standard matter 
Lagrangian. The massive spin-1 Lagrangian is \cite{deRham:2021efp}
\begin{equation}
    \mathcal{L}_{EPN} = - \frac{1}{4}F^{\mu \nu}F_{\mu \nu} + 
\Lambda^4\left(\mathcal{L}_0+\mathcal{L}_1 + \mathcal{L}_2 + \mathcal{L}_3 
\right),
\end{equation}
 where  
\begin{eqnarray}
&&	   \mathcal{L}_0 = \alpha_{0}(X) \label{ L0} \,,\\
&&	   \mathcal{L}_1 =  \alpha_{1}(X)  \mathcal{L}_1[\K] +  d_{1}(X) 
\frac{ \mathcal{L}_1[\nabla A]}{\Lambda^2} \label{ L1} \,,\\
&&	   \mathcal{L}_2 = \left[ \alpha_2(X) + d_2(X) \right] \frac{R}{\Lambda^2} 
+ \alpha_{2,X}(X)   \mathcal{L}_2[\K]  + d_{2,X}(X) \frac{  
\mathcal{L}_2[\nabla 
A]}{\Lambda^4} 
\label{ L2} \,,\\
&&	   \mathcal{L}_3 = \left[ \alpha_3(X) \K^{\mu \nu} +d_3(X) 
\frac{\nabla^{\mu} 
A^{\nu}}{\Lambda^2} \right] \frac{G_{\mu \nu}}{\Lambda^2}   - \frac16 
\alpha_{3,X}(X) 
  \mathcal{L}_3[\K] - \frac{1}{6} d_{3,X}(X) \frac{  \mathcal{L}_3[\nabla 
A]}{\Lambda^6} \label{ L3} 
\,,\ \
\end{eqnarray}
where $A_\mu$ is the vector field,  $\Lambda$ is an energy scale that   
  controls the strength of the vector self-interactions,  
 and the coefficients $\alpha_n(X)$ and $d_n(X)$  are   functions 
of $X=-\frac{1}{2\Lambda^2}\,A^{\mu}A_{\mu}$ (a subscript $X$ denotes 
differentiation with respect to $X$). Additionally,   
  $[\K] = \text{tr}(\K)$ 
is the trace of the tensor
  $\K\mupn = \X\mupn -\delta \mupn$ \cite{deRham:2010kj,deRham:2014zqa}, with 
$\X\mupn[A] = \left( \sqrt{\eta^{-1}f[A]}  \right)\mupn  $,  $\eta\mn$   the 
flat     Minkowski metric, and
\begin{equation}
	f\mn[A] = \eta\mn + 2 \frac{\p_{(\mu} A_{\nu)}}{\Lambda^2} + \frac{\p_{\mu} 
A^\rho \p_{\nu} A_\rho}{\Lambda^4} \,,
		\label{ deff2}
\end{equation}
 the     St\"uckelberg-inspired Lorentz 
tensor. Finally, the term  $G_{\mu\nu}$ in $\L_3$ is the Einstein tensor.
In summary,   massive Proca-Nuevo gravity is a consistent theory of a massive 
spin-1 fields coupled with gravity through minimal, non-minimal, and derivative 
terms.
 
Let us apply the above theory at a cosmological framework. In order to achieve 
this, we    focus on flat Friedmann-Robertson-Walker (FRW)    metric of the form
\begin{equation}
	\d s^2 = - \d t^2 + a^2(t) \delta_{ij} \d x^i \d x^j \,,
	\label{FRW}
\end{equation}
with $a(t)$ the scale factor, while for the vector field we assume  
\cite{deRham:2021efp}
\begin{equation} \label{ansatzphi}
	A_{\mu}\d x^{\mu} = -\phi(t)\d t \,,
\end{equation}
  with $\phi$ a scalar field.
  
Variation of the action with respect to the metric leads to the two Friedmann 
equations  \cite{deRham:2021efp}
\begin{eqnarray}
	&&H^2 = \frac{1}{3M_{Pl}^2} \left( \rho_{m} + 
\rho_{DE} \right) \,,
	\label{Fr1}\\
&&
	  \dot{H} + H^2 = - \frac{1}{6 M_{Pl}^2} 
\left(\rho_{m} + \rho_{DE} + 3 p_{m} + 3 
	p_{DE}\right) \label{Fr2}
\,,
\end{eqnarray}
while
variation with respect to $\phi(t)$ gives
\begin{equation}
	\alpha_{0,X} + 3 \left( \alpha_{1,X} + d_{1,X} \right) \frac{H 
\phi}{\Lambda^2} = 0 \,.
	\label{KleinGordon}
\end{equation}
In the above equations $\rho_{m}$ and $p_m$ are respectively the energy density 
and pressure of the perfect matter fluid, while we have introduce an effective 
dark energy sector with energy density and pressure given by
  \begin{eqnarray}
&&	\rho_{DE} \equiv \Lambda^4 \left[ - \alpha_{0} + \alpha_{0,X} 
\frac{\phi^2}{\Lambda^2} + 3 \left( \alpha_{1,X} + d_{1,X} \right) \frac{H 
\phi^3}{\Lambda^4} \right] \label{rhoDE} \,,
\\
&&	p_{DE} \equiv \Lambda^4 \left[ \alpha_{0} - \left( 
\alpha_{1,X} + d_{1,X} \right)  \frac{\phi^2 \dot{\phi}}{\Lambda^4} \right] \,.
	\label{ PExtPNhat}
\end{eqnarray}
What makes the present scenario interesting is that the  vector field equation, 
and thus due to the ansatz (\ref{ansatzphi}) the scalar field equation 
(\ref{KleinGordon}), is non-dynamical, i.e. it is just a constraint imposing an 
algebraic relation between $H$ and $\phi$. Hence, exactly due to the specific 
construction of action (\ref{action}), the resulting Friedmann equations depend 
only on the Hubble  function and not on the  vector condensate field. 
Therefore, the effective dark energy density and pressure are simply 
  \begin{eqnarray}
  \label{eq:DE-pres-rho}
    &&\rho_{DE} = \Lambda^4 \frac{c_m y^{2/3}}{2}\left( 
\frac{\Lambda^4}{M_{Pl}^2H^2} \right)^{1/3}\\
&& p_{DE} = 3M_{Pl}^2 H^2 \left(  - 1 - \frac{2\dot{H}}{3H^2} 
  \right) \,,
  \label{eq:DE-pres-rho2}
\end{eqnarray}
 with $c_m \equiv \frac{m^2 M_{\text{Pl}}^2}{\Lambda^4} \sim 1 $
  and $y = 4 \sqrt{\frac{6}{c_m}}$ \cite{deRham:2021efp}.  
  Finally, the equations close by considering the matter conservation equation 
$\dot{\rho}_m+3H(\rho_m+p_m)=0$, which according to the Friedmann equations 
(\ref{Fr1}),(\ref{Fr2}) leads to the dark energy conservation too, namely 
  $\dot{\rho}_{DE}+3H(\rho_{DE}+p_{DE})=0$.

In the following we use the redshift $z=-1+\frac{a_0}{a}$ as the independent 
variable, setting the present scale factor to $a_0=1$ (from now on, a 
subscript ``0'' denotes the value of a quantity at present time).
Furthermore, we focus 
on dust matter, namely we consider $p_m=0$, and thus the matter conservation 
equation gives $\rho_m=\rho_{m0}(1+z)^3$.
Additionally, as 
usual, we introduce the density parameters  
  \begin{eqnarray}
  \label{eqs:densities}
 &&  \Omega_{m}\equiv \frac{\rho_m}{3 M_{Pl}^2H^2}\\
   && \Omega_{DE}\equiv \frac{\rho_{DE}}{3 M_{Pl}^2H^2}.
\end{eqnarray}
 Thus, the first Friedmann equation (\ref{Fr1}) becomes  $\Omega_{m} 
+ \Omega_{DE} = 1$, which applied at present time, i.e. at $z=0$, and 
using (\ref{eq:DE-pres-rho}), gives  
\begin{equation}
    \frac{\Lambda^{16/3} c_m y^{2/3}
}{6 M_{Pl}^{8/3}H_0^{8/3}} = 1 - \Omega_{m0}.
\label{eq:ell}
\end{equation} 
Note that although the scenario has two intrinsic parameters, namely $\Lambda$ 
and $m$, the specific combination in which both of
them appear  can be eliminated in favour of just $\Omega_{m0}$.
Inserting (\ref{eq:ell})    back into 
(\ref{eq:DE-pres-rho}) we can eliminate $ 
c_m y^{2/3}$, as well as $\Lambda$, and therefore we obtain 
\begin{eqnarray}
  \label{eq:DE-pres-rhob}
    &&\rho_{DE} = 3    (1 - \Omega_{m0})
M_{Pl}^{2}H_0^{8/3}   
 H^{-2/3}.
\end{eqnarray}
Interestingly enough, in the scenario at hand the effective dark 
energy density depends only on the present matter density parameter 
$\Omega_{m0}$, as well as on the usual normalization factor, i.e. the present 
Hubble parameter $H_0$, and thus it has the same number of free parameters as 
$\Lambda$CDM cosmology, which for comparison we remind that it has
\begin{eqnarray}
  \label{rhoLCDM}
    &&\rho_{DE}|_{{}_{\Lambda CDM}} = 3    (1 - \Omega_{m0})
M_{Pl}^{2}H_0^{2}.
\end{eqnarray}
In summary, the  scenario of extended 
Proca-Nuevo cosmology, assuming a spatially flat
universe and considering only pressureless matter, has exactly the same number 
of parameters with 
$\Lambda$CDM cosmology, however its cosmological behavior will in general be 
different.

Inserting all these in the first  Friedmann equation (\ref{Fr1}), we obtain 
the 
simple algebraic equation 
\begin{equation}
    E(z)^2- \Omega_{m0}(1+z)^3- (1-\Omega_{m0})E(z)^{-2/3} = 0,
    \label{eq:hubble-rate-algebr}
\end{equation}
where $E(z)\equiv H(z)/H_0$ is the normalized Hubble function.  Furthermore, by 
aid of equations (\ref{eq:DE-pres-rho}),(\ref{eq:DE-pres-rho2}), we 
can define the effective equation-of-state parameter, namely  $w_{DE} \equiv 
p_{DE}/\rho_{DE}$, as
\begin{equation}
\label{eq:w_DE}
    w_{DE}(z) = \frac{E^{8/3}(z)}{1-\Omega_{m0}} \left[-1 + 
(1+z)\frac{2}{3}\frac{E'(z)}{E(z)}\right] .
\end{equation}

In summary, it is apparent 
that for large redshifts the term $E(z)^{-2/3}$ becomes much less important, 
and 
the model effectively reduces to pure general relativity plus cold dark 
matter scenario. On the other hand,  at late times, where $H\rightarrow 
H_0$ and thus $E(z)\rightarrow 1$, we observe 
that $\rho_{DE}$ behaves close to $\rho_{DE}|_{{}_{\Lambda CDM}}$ and for $z=0$ 
we obtain full coincidence. Hence, the scenario at hand of Proca-Nuevo 
cosmology has exactly the same number of parameters with 
$\Lambda$CDM cosmology, it reduces to the latter at early and present times, 
however at intermediate times it exhibits a non-trivial deviation. In the next 
section we desire to see whether this intermediate-time deviation can improve 
the fitting with the data comparing to 
$\Lambda$CDM paradigm.

\section{Observational Data and Analysis }
\label{Sec3}
In this Section  we review the   datasets that we will employ in our 
analysis, namely   data from Supernovae Ia (SNIa) observations and Cosmic 
Chronometers (CC)  direct measurements of the Hubble function. Additionally, we 
provide the various information criteria used for model comparison.  In 
what follows, $\phi^{\lambda}$ is the statistical vector, that contains the free 
parameters at hand.

\subsection{Supernovae Ia (SNIa)}

 One of the most extensively studied class of standard candles in cosmology are 
the Supernovae Type Ia (SNIa). We incorporate the  binned Pantheon sample   
\cite{scolnic2018complete}. The entire dataset is   approximated by the binned 
dataset 
comprising $N = 40$ data points within the redshift range $0.01 \lesssim z 
\lesssim 1.6$.

The chi-square function for this dataset is expressed as:
  \begin{equation}
          \chi^{2}_{SN Ia}\left(\phi^{\nu}\right)={\bf \mu}_{\text{SNIa}}\,
          {\bf C}_{\text{SNIa},\text{cov}}^{-1}\,{\bf \mu}_{\text{SNIa}}^{T}\,,
          \end{equation}
with    ${\bf 
\mu}_{\text{\text{SNIa}}}=\{\mu_{1}-\mu_{\text{th}}(z_{1},\phi^{\nu})\,,\, 
...\,, \,   \mu_{N}-\mu_{\text{th}}(z_{N},\phi^{\nu})\}$. 
  The distance modulus is defined as $\mu_{i} = \mu_{B,i}-\mathcal{M}$, with 
$\mu_{B,i}$ the apparent magnitude at maximum in the rest frame for redshift 
$z_{i}$.  The introduction of the free parameter 
$\mathcal{M}$ is essential due to the dependence of the observable distance 
modulus, $\mu_{obs}$, on assumptions related to $H_0$ and the fiducial 
cosmology. By absorbing artifacts originating from the transformation of flux 
observations to distance modulus estimates, $\mathcal{M}$ enables independence 
from the fiducial cosmology, as discussed in \cite{scolnic2018complete}.  
Additionally, the theoretical 
form of the distance modulus is given by:
\begin{equation}
\mu_{\text{th}} = 5\log\left[\frac{d_{L}(z)}{\text{Mpc}}\right] + 25,
\end{equation}
with $
d_L(z) = c(1+z)\int_{0}^{z}\frac{dx}{H(x,\phi^{\nu})}
$
 the luminosity distance for spatially flat   geometry. Note that $\mathcal{M}$ 
and the normalized Hubble constant $h$ exhibit intrinsic degeneracy within the 
context of the Pantheon dataset. 

\subsection{Cosmic Chronometers (CC)}

We utilize data from the most recent compilation of the $H(z)$ 
dataset, as provided by \cite{magana2018cardassian}. We focus on data 
obtained from cosmic chronometers (CC), which represent massive galaxies 
evolving at a relatively slow pace during distinct intervals of cosmic time. By 
leveraging their differential age, it becomes possible to directly measure the 
Hubble rate \cite{michele2022unveiling}. A notable advantage of utilizing the 
differential age of passively evolving galaxies is that the resulting Hubble 
rate measurement rely on minimal assumptions about the underlying 
cosmology, i.e. FRW geometry. Moreover, we exclude CC measurements from 
\cite{simon2005constraints} in accordance with Ref. \cite{ahlstrom2023use}, 
which found that they are not reproducible. Our analysis incorporates a total 
of 
$N=22$ measurements of the 
Hubble expansion, covering the redshift range $0.07 \lesssim z \lesssim 2.0$.

In this context, the corresponding $\chi^2_{H}$ function is expressed as:
  \begin{equation}
          \chi^{2}_{H}\left(\phi^{\nu}\right)={\bf \cal H}\,
          {\bf C}_{H,\text{cov}}^{-1}\,{\bf \cal H}^{T}\,,
          \end{equation}
with ${\bf \cal H 
}=\{H_{1}-H_{0}E(z_{1},\phi^{\nu})\,,\,...\,,\,
          H_{N}-H_{0}E(z_{N},\phi^{\nu})\}$ and $H_{i}$ are the observed Hubble 
rates at  redshift $z_{i}$ ($i=1,...,N$). 

     \subsection{Likelihood analysis and model selection}

Given $P$ independent observational data-sets and assuming Gaussian errors, 
 the total likelihood function is defined as:
 \begin{equation}
        \mathcal{L}_{\text{tot}}(\phi^{\psi}) \sim \prod_{p=1}^{\mathcal{P}} 
\exp[-\chi^2_{p}(\phi^{\psi})]\,,
        \end{equation}
where   the corresponding expression for $\chi_{\text{tot}}^2$ is written as
       \begin{equation}
        \chi_{\text{tot}}^2 = \sum_{p=1}^{\mathcal{P}}\chi^2_{p}\,.
        \end{equation}
The statistical vector has a dimension of $k$, comprising $\nu$ parameters of 
the model under consideration plus $\nu_{\text{hyp}}$ hyper-parameters from the 
utilized data sets, yielding   $k = \nu + \nu_{\text{hyp}}$. 
In the above expression  $\phi^{\mu}$ is the vector 
containing the free parameters, which in our case is
$\phi^{\mu} = 
\{\Omega_{m0},h,\mathcal{M} \}$ and $\mathcal{P} = \{CC, SNIa\}$.  Lastly, to 
obtain the posterior distributions of the model parameters, given the data, we 
use a Markov Chain Monte Carlo (MCMC) sampler. Instead of the standard 
Metropolis - Hastings algorithm, to avoid the need of fine-tuning for the 
hyper-parameters, we use an affine-invariant Markov Chain Monte Carlo sampler 
as implemented within the open-source Python package emcee 
\cite{foreman2013emcee,foreman2019emcee}, involving 1000 chains (walkers) and 
2500 steps 
(states). Regarding the convergence of the MCMC algorithm, we use the 
traditional Gelman-Rubin criterion and also the auto-correlation time analysis.

In the assessment of cosmological models based on their   predictions 
with respect to the available data, we employ three widely recognized criteria: 
the Akaike Information Criterion (AIC), the Bayesian 
Information Criterion (BIC), and the Deviance Information 
Criterion (\cite{liddle2007information} and references therein).

The AIC criterion addresses the issue of model adequacy from an information 
theory perspective. Specifically, it serves as an estimator of the 
Kullback-Leibler information and possesses the property of asymptotic 
unbiasedness. Under the standard assumption of Gaussian errors, the AIC 
estimator is expressed as \cite{liddle2007information}:
\begin{equation}
\text{AIC}=-2\ln(\mathcal{L}{\text{max}})+2k+\frac{2k(k+1)}{N{\rm tot}-k-1}.
\end{equation}
Here, $\mathcal{L}{\text{max}}$ represents the maximum likelihood of the 
considered data set(s), and $N{\rm tot}$ is the total number of data points. 
For   large   $N_{\rm tot}$  the expression simplifies to 
$\text{AIC}\simeq -2\ln(\mathcal{L}_{\text{max}})+2k$, which corresponds to the 
standard form of the AIC criterion. Consequently, it is advisable to utilize 
the 
modified AIC criterion in all cases \cite{liddle2007information}.

The Bayesian Information Criterion (BIC) serves as an estimator of the Bayesian 
evidence, and its expression is given by:
\begin{equation}
\text{BIC} = -2\ln(\mathcal{L}{\text{max}}) + k \cdot 
\text{log}(N{\text{tot}}).
\end{equation}
On the other hand, the Deviance Information Criterion (DIC) is formulated by 
incorporating concepts from both Bayesian statistics and information theory 
\cite{liddle2007information}. It is expressed as:
\begin{equation}
\text{DIC} = D(\overline{\phi^\mu}) + 2C_{B}, 
\end{equation}
with   $C_{B} = \overline{D(\phi^\mu)} - D(\overline{\phi^\mu})$ the Bayesian 
complexity, where the overline represents the   
mean value. Moreover, $D(\phi^\mu)$ corresponds to the Bayesian Deviation, 
which, for a general class of distributions (the exponential family), is given 
by $D(\phi^\mu) = -2\ln(\mathcal{L(\phi^\mu)})$. This   is closely tied 
to the effective degrees of freedom \cite{liddle2007information}, representing 
the 
number of parameters that actually contribute to the fitting.

In the task of ranking a set of competing models based on their fitting quality 
to observational data, we employ the  aforementioned criteria, 
specifically focusing on the relative difference in Information Criterion (IC) 
values within the given model set. The difference $\Delta 
\text{IC}{\text{model}} = \text{IC}{\text{model}} - \text{IC}_{\text{min}}$, 
compares each model's IC value to the minimum IC value in the set of competing 
models.
We use the rule
\begin{equation}
\label{prob_per_model}
P \simeq \frac{e^{-\Delta \text{IC}_{i}}}{\sum_{i=1}^{n}e^{-\Delta 
\text{IC}_{i}} },
\end{equation}
in order to  assess the ``degree of belief'' for each model  
\cite{burnham2004multimodel}, where
 the index $i$ runs over the set of $n$ models. 
 Finally, according to Jeffreys scale  \cite{Kass:1995loi}, 
when $\Delta\text{IC}\leq 2$ the model is statistically compatible with the 
most favored model by the data, when  $2<\Delta\text{IC}<6$ implies 
a moderate tension between the two models, while   the case 
$\Delta\text{IC}\geq 10$ indicates a significant tension.

\section{Results and Discussion}
\label{Sec4}

\begin{figure}[ht]
\includegraphics[scale=0.73]{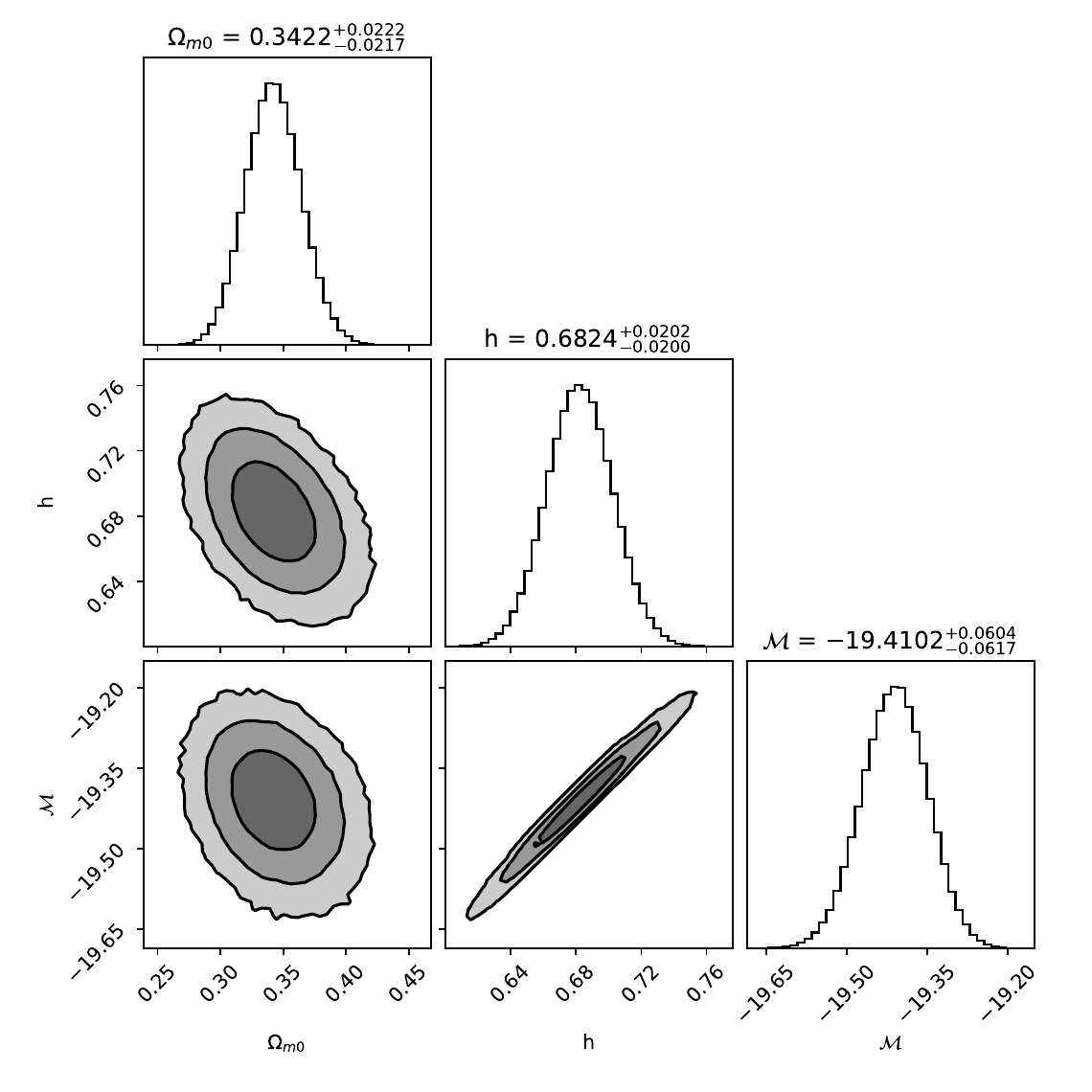} 
\caption{{\it{
The $1\sigma$, $2\sigma$ and $3\sigma$ iso-likelihood contours for  the
extended Proca-Nuevo (EPN) cosmology, for the various  two-dimensional subsets 
of 
the parameter space $(\Omega_{m0},h,\mathcal{M})$, using  the joint 
analysis of SNIa/CC  datasets. Furthermore, we present the 
 mean values of the parameters. }}} 
 \label{fig:posterior-likelihood}
\end{figure}

We have now all the machinery needed to proceed to the observational 
confrontation of extended Proca-Nuevo  gravity at cosmological level. We 
perform the 
analysis described in the previous section, and we summarize the results  for 
the posterior parameter values
 in Table \ref{tab:Results1}, while in Table \ref{tab:Results2} we present the 
values for model selection criteria. The posterior distributions of the model 
parameters are presented in Fig \ref{fig:posterior-likelihood} as 
iso-likelihood contours on two-dimensional sub-spaces of the parameter space 
(triangle plot).

\begin{table}[ht]
\begin{center}
\begin{tabular}{ccccccc} \hline \hline
Model & $\Omega_{m0}$ & $h$ &  $\mathcal{M}$ & $\chi_{\text{min}}^{2} $&  
$\chi_{\text{min}}/dof$  
\vspace{0.05cm}\\ 
\hline
\hline
   EPN cosmology & $0.342 \pm 0.022$      & $0.682 \pm 0.022$  &$-19.41 \pm 
0.06 
$ 

 & $ 
45.42
$ 
& $0.77$\\ 
    
$\Lambda$CDM & $ 0.300 \pm 0.022$    & 
$0.690_{-0.018}^{+0.019} $ & $-19.38 \pm 0.05$ & $53.94$ & $0.79$  \\  
\hline\hline
\end{tabular}
\caption[]{Observational constraints and the
corresponding $\chi^{2}_{\rm min}$ for 
 the Extended Proca-Nuevo (EPN) cosmology, as well as for 
$\Lambda$CDM paradigm,  using  the joint 
analysis of SNIa/CC  datasets.   For 
direct comparison we additionally include the concordance $\Lambda$CDM 
scenario.}
\label{tab:Results1}
\end{center}
\end{table}

\begin{table}[ht]
\begin{center}
\tabcolsep 4.0pt
\vspace{1mm}
\begin{tabular}{ccccccc} \hline \hline
Model & AIC & $\Delta$AIC & BIC &$\Delta$BIC & DIC & $\Delta$DIC
 \vspace{0.05cm}\\ \hline
\hline
EPN cosmology & 51.8 & 0.0 & 57.8  & 0.0 & 51.3 & 0.0 \\  
$\Lambda$CDM & 60.3  & 8.5 & 66.7 & 8.9 & 59.9  & 8.6  \\ 
\hline\hline
\end{tabular}
\caption{The information criteria 
AIC, BIC and DIC, for the Extended Proca-Nuevo (EPN) cosmology, as well as for 
$\Lambda$CDM paradigm, alongside
the relative difference from the best-fitted model $\Delta\text{IC} 
\equiv \text{IC} - \text{IC}_{\text{min}}$.
\label{tab:Results2}}
\end{center}
\end{table}

\begin{figure}[ht]
\centering
\includegraphics[scale=0.73]{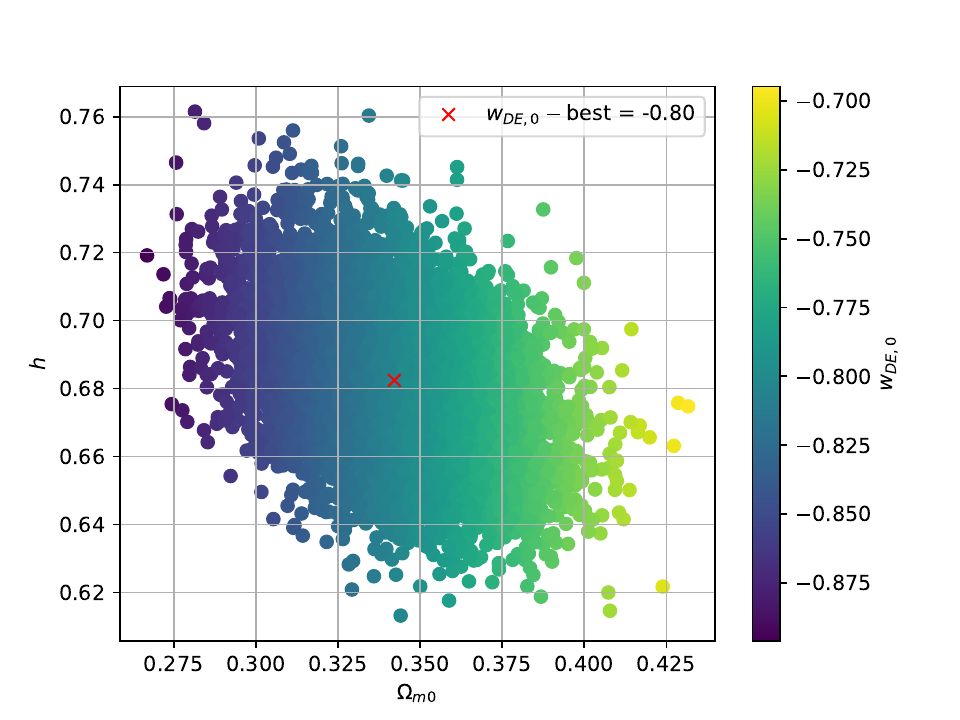} \vspace{-0.2cm}
\hspace{-1.cm}
\includegraphics[scale=0.42]{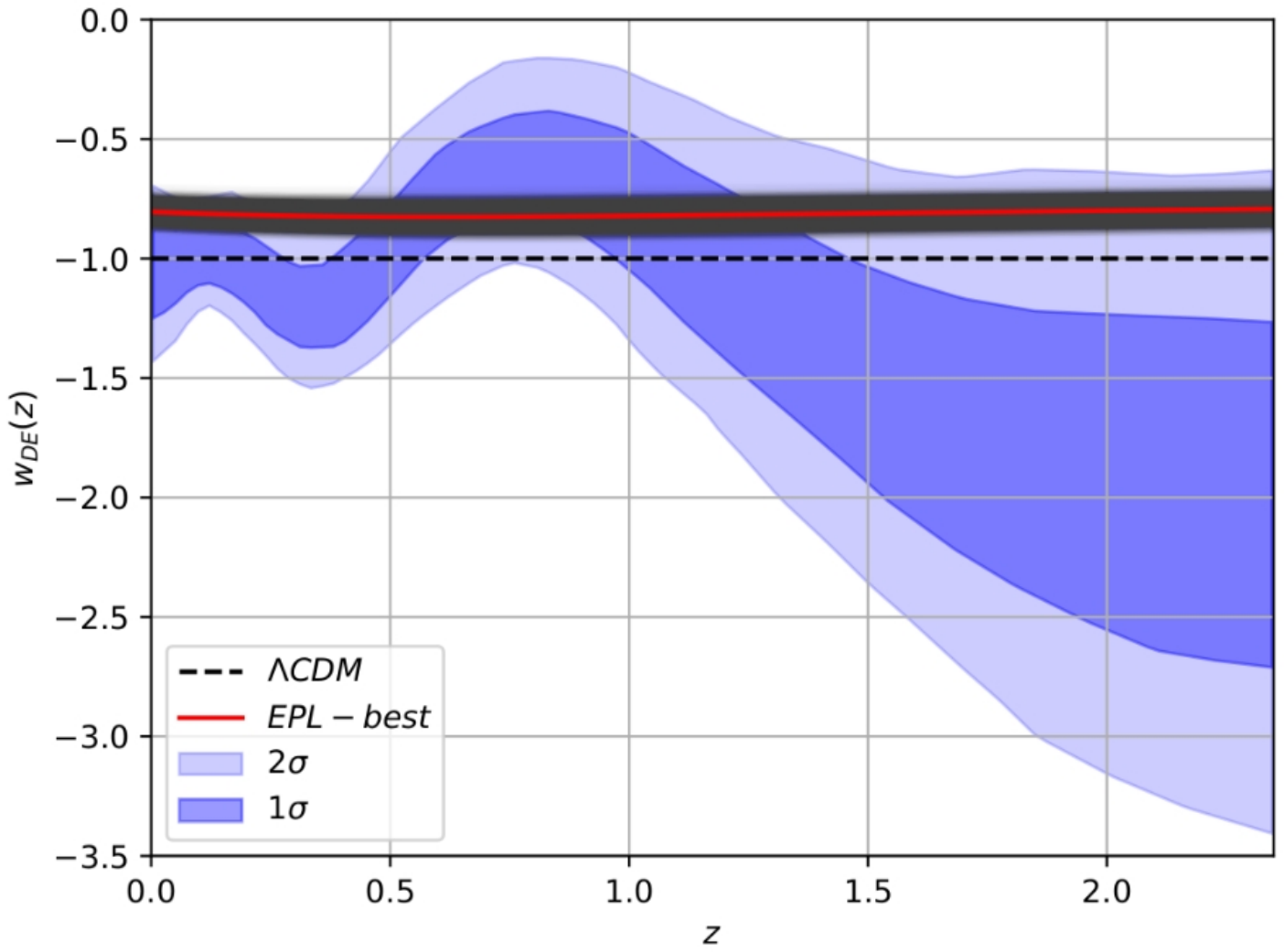} 
\caption{{\it{ \textbf{Upper panel:}
Scatter plot of 8500 re-sampled points from the posterior parameters 
distributions with regard to the present value of the equation-of-state 
parameter $w_{DE}(z=0)$ for  the
extended Proca-Nuevo (EPN) cosmology. The point which corresponds to the most 
probable value of the parameters is given as red cross.
\textbf{Lower panel:} Reconstruction of the equation-of-state parameter 
\eqref{eq:w_DE} for  EPN cosmology. Additionally, the blue, shaded areas, 
correspond to $1\sigma$ (deep blue) and $2\sigma$ (light blue) regions for the 
model independent reconstruction of $w_{DE}(z)$ taken from 
\cite{zhao2017dynamical}   using similar datasets. The 
red continuous curve corresponds to the most probable parameter values for the 
EPN scenario and the black dashed line to the concordance value of $w_{DE} = 
-1$.}}} 
\label{fig:wde}
\end{figure}

As we observe, the scenario at hand is in agreement with observations. 
Nevertheless, the most interesting result is that, by means of model selection 
criteria, for the considered datasets the Extended Proca-Nuevo cosmology is 
preferred over the concordance one. In particular, all considered model 
selection criteria (AIC, BIC, DIC) suggest the above result, with the 
difference 
of  $\Delta IC$    being $\sim 8$ in all cases (see Tab. \ref{tab:Results2}). 
To 
better understand these $\Delta IC$ values, we calculate the ``degree of 
belief'' \eqref{prob_per_model}  given the data for extended Proca-Nuevo 
scenario  which is $P \sim 0.99998$, while the corresponding one for 
$\Lambda CDM$ 
is $P \sim 1e-6$. 

In comparison with other cosmological models that potentially manage to 
perform better than $\Lambda CDM$, such as Running Vacuum models (RVMs) 
\cite{rezaei2022cosmographic} ($\Delta IC_{\textrm{max}} \leq 3$), or $f(Q)$ 
cosmologies \cite{atayde2023f} ($\Delta IC_{\textrm{max}} \sim 3$) and 
\cite{Anagnostopoulos:2021ydo} ($\Delta IC_{\textrm{max}} \sim 0.3$),  the 
extended Proca-Nuevo cosmology shows the best performance against the 
concordance cosmology. One might be inclined to interpret the latter findings 
directly. However, we would desire not to make such a strong and definitive 
statement, since for the moment we have focused solely on background data 
for the late universe, not including perturbation analysis, namely utilization 
of $f\sigma8$ data as it was done in \cite{Anagnostopoulos:2021ydo}, and/or CMB 
data. 
 
On the other hand, in terms of our results for the  posterior parameter values, 
we note an approximately 10\% increase in the $\Omega_{m0}$ parameter compared 
to the value in the concordance model, accompanied by a subsequent reduction in 
the Hubble constant $h$. These values for the extended Proca-Nuevo cosmology 
lie 
within approximately 2$\sigma$ of PLANCK  results, namely $\Omega_{m0} = 
0.315 \pm 0.007$ and $h = 0.674 \pm 0.005$ \cite{Aghanim:2018eyx}.

Moreover, we use the data and expression \eqref{eq:w_DE} and we  reconstruct 
the 
 equation-of-state parameter of dark energy \eqref{eq:w_DE}, depicting it in 
Fig. \ref{fig:wde}. In particular, in the upper panel of Fig.  \ref{fig:wde} we 
illustrate the present-day value of the equation-of-state parameter, 
$w_{DE}(z=0) \equiv w_{DE,0}$, for various parameter vectors. It is observed 
that as we approach the concordance model with $w_{DE,0} = -1$, smaller values 
of $\Omega_{m0}$ are attained. In the lower panel of  Fig. \ref{fig:wde}, we 
illustrate the reconstructed evolution of $w_{DE}(z)$ for extended Proca-Nuevo 
cosmology scenario (red line), using a re-sampling of the posterior 
parameter distribution. For the benefit of the reader, we additionally include 
the $\Lambda$CDM value $w_{DE}(z) = -1$  (dashed black line) as well as the 
model independent reconstruction of $w_{DE}(z)$ taken from 
\cite{zhao2017dynamical} that has been obtained using similar datasets. We 
observe statistical compatibility for $\sim 1 $ to $ 2 \sigma$ levels, between 
our reconstructed $w_{DE}(z)_{EPN}$ and the model independent one from 
\cite{zhao2017dynamical}, which serves as an additional check for the 
correctness of our analysis.

Furthermore, the increased  value at present time $w_{DE}(z=0)_{EPN}$ in 
comparison with the corresponding value  for $\Lambda CDM$ scenario is related 
to the 
reduced $h$ and increased $\Omega_{m0}$. Both $\Omega_{m0}$ increment and $H_0$ 
decrement may cause changes regarding the matter clustering, however again we 
must restrain ourselves until the full analysis is performed. According to 
\cite{heisenberg2023simultaneously}, for a cosmology which asymptotically 
reaches $\Lambda CDM$ at large redshifts, in order to  solve $H_0$  tension, 
while having constant $G(z) \equiv G_{N}$,  a phantom crossing is needed, which 
according to the lower panel of Fig. \ref{fig:wde}  does not seem to be 
realized. However, this could not be a problem, as due to the included spin-1 
field of EPN gravity, emergent interactions at the perturbative level provide 
scale-dependent effects and significant modifications on the friction term on 
the matter overdensity $\delta_{m}$ equation via the modified sound velocity 
\cite{deRham:2021efp}, 
thus violating part of the assumptions used by 
\cite{heisenberg2023simultaneously}, e.g the incorporation of new physics 
effects only via $\Delta G \equiv G_{eff}(z) - G_{N}$.

\section{Conclusions}
\label{Conclusions}

In this work we confronted massive Proca-Nuevo  gravity with cosmological 
observations. The former is a recently proposed non-linear theory inspired by 
dRGT massive gravity, but involving a massive spin-1 field, that can be 
extended 
incorporating operators of the Generalized Proca class without compromising 
the essential primary constraint crucial for consistency. Then the theory can 
be coupled to gravity and can be covariantized in a way that exhibits 
consistent and ghost-free cosmological solutions. Specifically, one can recover 
the thermal history of the universe, and obtain a late-time accelerated epoch. 
Additionally, these cosmological solutions are well-behaved at the perturbative 
level, without experiencing instabilities and superluminalities, which was the 
typical problem of the basic massive gravity scenarios.

When it is applied at a cosmological framework, massive Proca-Nuevo gravity 
induces extra terms in the Friedmann equations that can be collectively 
absorbed into an effective dark-energy sector. The interesting feature of the 
special non-linear construction is that the Klein-Gordon equation that arises 
form the scalar field parametrizing the vector ansatz, is  non-dynamical, 
namely 
it is just a constraint imposing an algebraic relation between the field and 
the 
Hubble function. This allows to eliminate the field in favor of the Hubble 
function, and thus the effective dark energy sector finally depends only on the 
Hubble function. In 
summary, the cosmological scenario of massive Proca-Nuevo gravity is different 
than $\Lambda$CDM paradigm, but possessing exactly the same number of free 
parameters.

We used data from Supernovae Ia (SNIa) and Cosmic Chronometers (CC) 
observations, in order to extract constraints of the free parameters of the 
theory. In particular, we provided the corresponding likelihood-contours, as 
well as the best-fit values and the 1$\sigma$ intervals for the parameters, 
showing that the scenario is in agreement with observations. Nevertheless, 
interestingly enough, application of various information criteria showed that 
 massive Proca-Nuevo gravity can be more efficient comparing to $\Lambda$CDM 
concordance model in fitting the data. The reason for this is that the scenario 
does include a dynamical dark energy, but without extra parameters comparing to 
$\Lambda$CDM  model, and hence  it can lead to an improved behavior.
Finally, the  reconstructed    dark-energy  equation-of-state parameter  
showed statistical compatibility at $\sim 1 $ to $ 2 \sigma$ levels, 
with the model-independent, data-driven reconstructed one.

In summary, massive Proca-Nuevo gravity and cosmology may potentially challenge 
$\Lambda$CDM paradigm. However, there are definitely additional investigations 
that one must perform in order to reach to such a conclusion. Among these, is 
the confrontation with observations at the perturbative level, using data from 
Large Scale Structure (i.e. $f\sigma_8$ data) and other probes. Such a study 
lies beyond the 
present work and it is left for a future project.

\begin{acknowledgments}
 
The authors   acknowledge the contribution of the LISA CosWG, and of   COST 
Actions  CA18108  ``Quantum Gravity Phenomenology in the multi-messenger 
approach''  and  CA21136 ``Addressing observational tensions in cosmology with 
systematics and fundamental physics (CosmoVerse)''.

\end{acknowledgments}

 \bibliographystyle{utphys}
\bibliography{references}

\end{document}